\documentclass[%
amsmath,amssymb,
aps,pra,twocolumn,10pt
]{revtex4-2}

\usepackage{graphicx}% Include figure files
\usepackage{dcolumn}% Align table columns on decimal point
\usepackage[normalem]{ulem} 
\usepackage{bm}% bold math
\usepackage{braket}
\usepackage{hyperref}% add hypertext capabilities 
\usepackage{soul}
\usepackage{color}
\usepackage{tikz}
\usetikzlibrary{shapes}
\usepackage{wasysym}

\hypersetup{
    colorlinks,
    linkcolor={red!50!black},
    citecolor={blue!50!black},
    urlcolor={blue!80!black}
}

\begin{document}

\title{Quench dynamics in topologically non-trivial quantum many-body systems}

\author{Sarika Sasidharan Nair}
\email{sarika.sasidharannair@oist.jp}
\affiliation{Quantum Systems Unit, Okinawa Institute of Science and Technology\\ Graduate University, Onna-son, Okinawa 904-0495, Japan }

\author{Giedrius \v{Z}labys}
\affiliation{Quantum Systems Unit, Okinawa Institute of Science and Technology\\ Graduate University, Onna-son, Okinawa 904-0495, Japan }

\author{Wen-Bin He}
\affiliation{Quantum Systems Unit, Okinawa Institute of Science and Technology\\ Graduate University, Onna-son, Okinawa 904-0495, Japan }

\author{Thom\'as Fogarty}
\affiliation{Quantum Systems Unit, Okinawa Institute of Science and Technology\\ Graduate University, Onna-son, Okinawa 904-0495, Japan }

\author{Thomas Busch}
\email{thomas.busch@oist.jp}
\affiliation{Quantum Systems Unit, Okinawa Institute of Science and Technology\\ Graduate University, Onna-son, Okinawa 904-0495, Japan }

\date{\today}

\begin{abstract}
We investigate the non-equilibrium dynamics of a ground state fermionic many-body gas subjected to a quench between parameter regimes of a topologically non-trivial Hamiltonian. By focusing on the role of the chiral edge states inherent to the system, we calculate the many-body overlap and show that the characteristic monotonic decay of the orthogonality catastrophe with increasing system size is notably altered. Specifically, we demonstrate that the dynamics are governed not solely by the total particle number but rather by the number of occupied single-particle edge states. This behavior is further explained through an analysis of the full work probability distribution, providing a deeper understanding of the system’s dynamics.
\end{abstract}

\maketitle

\section{\label{sec:Intro}Introduction}

Topological properties of quantum many-body systems have been a center of interest ever since the discovery of the Quantum Hall effect \cite{vonKlitzing:80,Tsui:82,Thouless:82}. In the recent decade ultracold atomic systems have become powerful quantum simulators to study topologically non-trivial quantum matter \cite{atala2013,aidelsburger2015,goldman2016}, as they are clean and highly controllable. This allows one to isolate specific settings and study the relevant topological properties in more detail than in conventional solid-state systems \cite{hsieh2009,hasan2010}. In particular, as one of the fundamental characteristics of topological matter is a robustness against local disorder, it is interesting to study the response of such a system to non-adiabatic perturbations. Over the past couple of years this topic has been addressed in several theoretical \cite{dora2011,perfetto2013,barnett2013,patel2013} and experimental \cite{sun2018,dalessio2015,caio2015,kells2014,degottardi2011,zhang2018} studies, with most of these considering quenching between different topological phases or between trivial and nontrivial phases. 

In this work we study the effect of the presence of chiral edge states on the non-equilibrium dynamics of a quantum many-body system of ultracold atoms. For this we consider a fundamental model where a number of fermions are trapped in a finite sized lattice inside a box potential in a quasi-one-dimensional setting. Assuming that the barriers between different lattice sites are given by point-like $\delta$-functions \cite{lkacki2016,wang2018}, and allowing for all individual barrier positions and strengths to have independent values, this is the so-called Arbitrary Finite Kronig-Penney (AFKP) model, for which one can find the exact spectrum using the Bethe ansatz \cite{Reshodko:19}. The fact that the barrier positions and strengths in AFKP can be freely chosen, allows one to construct a large number of interesting settings ranging from  perfect periodicity, to localised disorder and to complete randomness. In this work we consider a perfectly periodic arrangement of barriers of identical height, but keep the position of this lattice with respect to be external box potential as a degree of freedom. The shift between the left hand side edge of the box and the left-most barrier potential then provides a quasi-dimension, which allows the system to become topologically non-trivial and support chiral edge states \cite{Reshodko:19}.  

We will investigate the effects of the chiral edge states on non-equilibrium dynamics that originates when quenching the shift parameter. While non-equilibrium dynamics can be very complex in many-particle systems, it can be dealt with in systems of non-interacting fermions if the single particle eigenstates are known. Here, by calculating static and dynamic overlaps between the many-body states before and after the quench, we show that the presence of edge states has a significant influence on the non-equilibrium dynamics and, in particular, we find that the presence of an edge mode leads to a deviation from the characteristic monotonically decaying behaviour of the orthogonality catastrophe (OC) \cite{anderson1967}. The detailed dynamics can be explained by identifying the main excitations using the work probability distribution, which allows us to show that the system response to the quench is governed by the number of occupied states around the highest occupied edge state and therefore group the dynamics of many-body states with different  particles numbers. While the model we describe makes certain approximations, it is  worth noting that recent theoretical and experimental progress has allowed to make nano-scale potentials for ultracold atomic systems that approximate delta-function barriers to a very high degree \cite{wang2018,lkacki2016,gediminas2021optical}. 

The structure of the manuscript is as follows. In Section \ref{Sec:Model}, we introduce the model, provide some background on quantum quenches and briefly describe the survival probability which quantifies the quenches in the system.  In Section \ref{sec:Fidelity}, we explore the effects of the chirality of the single particle edge states on the many-body quench dynamics by looking at the static and dynamical overlap, and the corresponding survival probability. We explain how the edge states affect the transition probabilities between different ground states and how this leads to a specific behaviour of the orthogonality catastrophe. We finally calculate the work probability distribution and show how the presence of edge states affects the excitation of particles after the many-body quench. We conclude with a brief summary and discussion in Section \ref{Sec:conclusions}.

\section{Model}
\label{Sec:Model}

We consider a gas of ultracold, non-interacting fermionic atoms  at zero temperature in  a one-dimensional trap. The gas is confined in an infinite box potential of length $L$ and inside the box $M$ point-like Dirac-delta scatterers of identical height, $h$, are distributed at regular positions $(y_{1},...,y_{M})$. The positions of the scatterers are ordered from left to right. Such a finite system represents one realisation of the Arbitrary Finite Kronig-Penney (AFKP) Hamiltonian \cite{Reshodko:19}, which is of the form
  \begin{equation}
    \mathcal{H}(x) = - \frac{1}{2} \frac{d^2}{dx^2} + V(x)\;,
  \label{eq:Hmltn}
  \end{equation}
  and where the potential is described by
  \begin{equation}
    V(x)=
    \begin{cases}
        \sum_{n=1}^{M}h_{n}\delta(x-y_{n}), & \text{for $-\frac{L}{2}<x<\frac{L}{2}$}\;,\\
        \infty, & \text{otherwise}\;.
    \end{cases}
    \label{eq:potn}
\end{equation} 
For notational simplicity, we have set the reduced Planck constant and mass to be unity throughout, $\hbar=m=1$, while the length is in units of the lattice constant $a=L/M$ and therefore the energy is in units of $a^{-2}$.

\begin{figure}[tb]
    \centering
    \includegraphics[width=1\columnwidth]
    {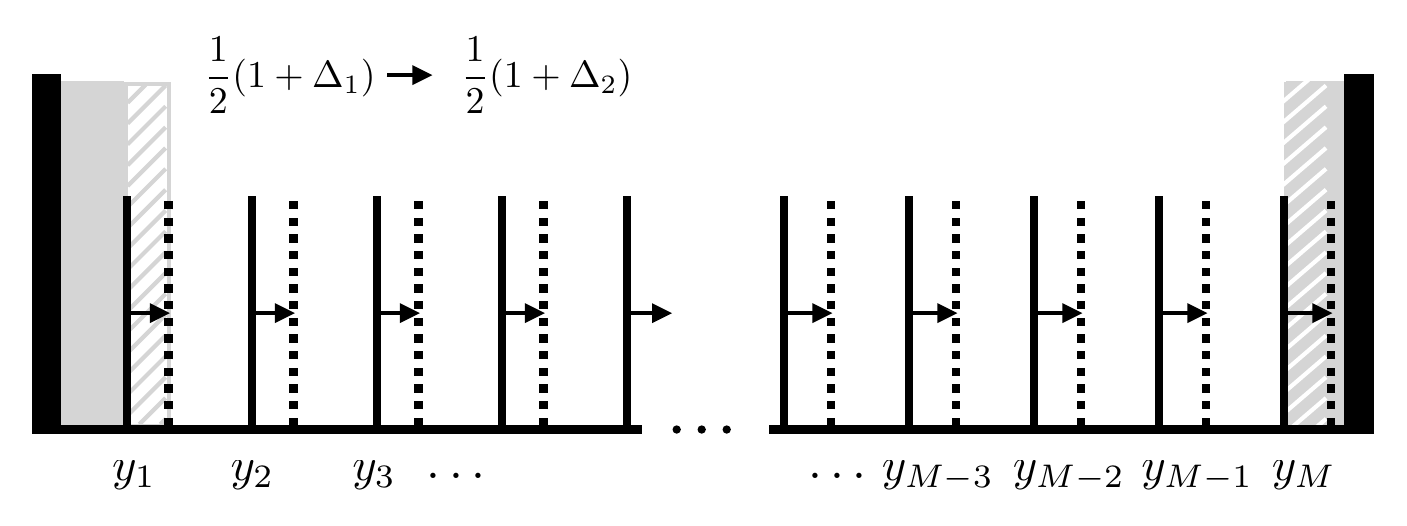}
    \caption{Schematic of the AFKP potential we consider for two different values of the shift parameter $\Delta$. The 10 barriers are placed at the positions of $(y_1,...,y_M)$ and form an equidistantly spaced lattice of constant height $h$. A quench is initialised by shifting the lattice with respect to the walls of the box and quantified by the shift parameter $\Delta_1\rightarrow\Delta_2$.
    } 
    \label{fig:Schematic}
\end{figure}

\begin{figure}[tb]
    \centering
    \includegraphics[width=\columnwidth]
   {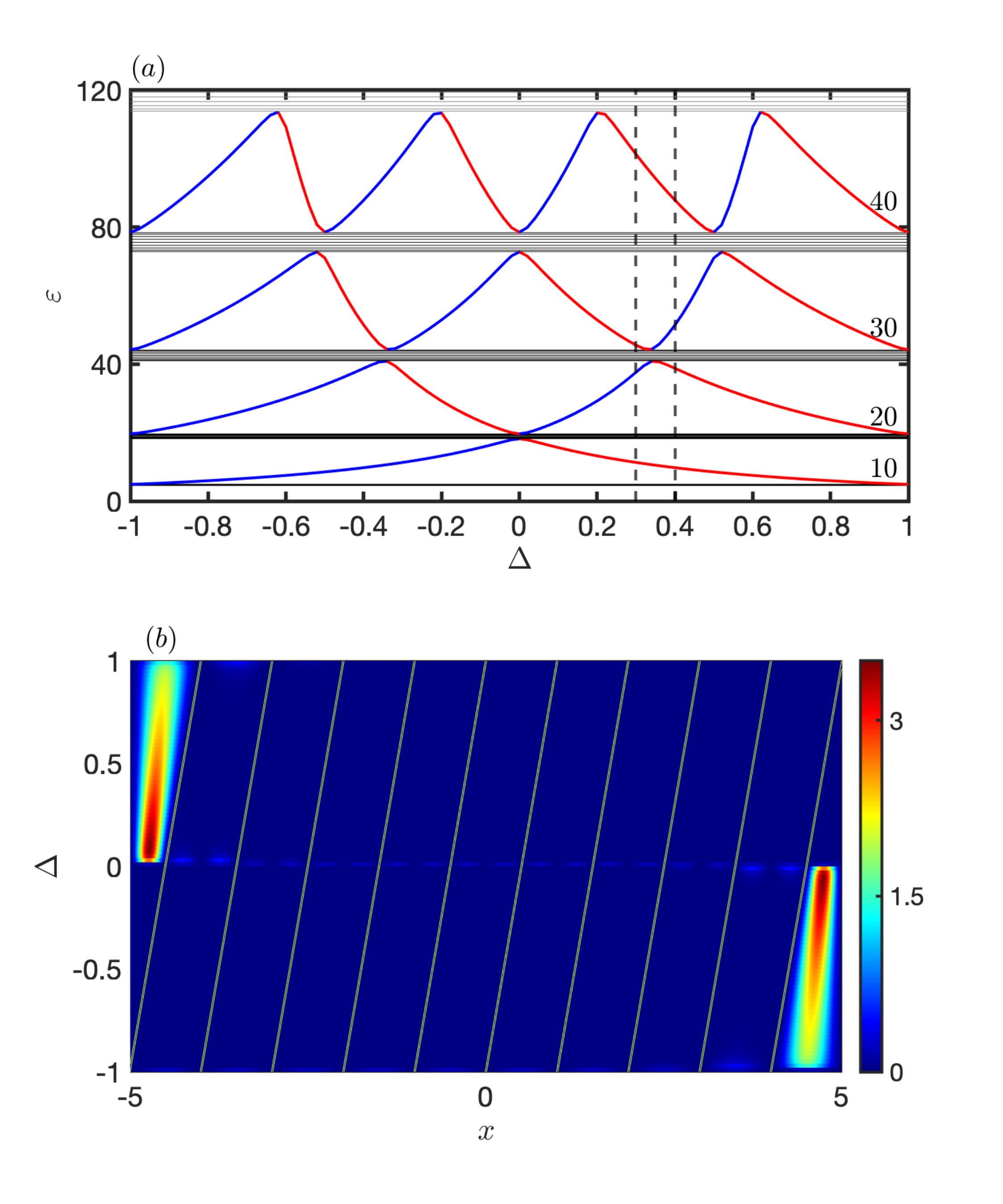}
    \caption{(a) Energy spectrum of the AFKP Hamiltonian with 10 equidistant barriers of height $h=50$  in a box of length $L=10$ as a function of the shift $\Delta$. The black lines corresponds to band-states that are delocalised over the whole system, whereas for the gap states the colour indicates the chirality, with blue indicating a localisation on the right, and red a localisation on the left. The numbers on the right hand side correspond to the quantum number of each gap eigenstate. The dashed vertical lines at $\Delta=0.3$ and $\Delta=0.4$ are a guide to the eye to identify the chirality of the edge states for these values of $\Delta$. 
    (b) Density of the 10$^\text{th}$ eigenstate, which is the lowest lying edge state. Yellow lines indicates the position of barriers within the box at each values of $\Delta$.
    }
    \label{fig:EnergySpectrumTopologicalState}
\end{figure}

Systems with one-dimensional periodic potentials, that are confined in a potential with open boundary conditions, are known to be able to host topological edge states under global  potential translation \cite{lang2012,zheng2014}. In our model this can be realised by assuming the equidistantly spaced lattice of equal-height $h$ barriers is shifted inside the box potential with respect to the box edge (see Fig.~\ref{fig:Schematic}). The positions of the barriers then depend on the value of the shift parameter $\Delta$ and can be parametrised by
\begin{equation}
    y_{n}=-\frac{L}{2}+\left(n+\frac{\Delta-1}{2}\right)\frac{L}{M}\;,
    \label{eq:position}
\end{equation} 
where $\Delta \in{[-1,1]}$ and for $\Delta=-1$ the left-most barrier coincides with the left wall, while for $\Delta=1$ the rightmost barrier coincides with the right wall. 
For the remainder of the work, and without loss of generality, we will focus on a system of trapped particles in a box of length $L=10$ which contains $M=10$ barriers and the height of the barrier set to $h=50$. The single-particle eigenspectrum of such a system  as a function of $\Delta$ is shown in Fig.~\ref{fig:EnergySpectrumTopologicalState}(a) and, in addition to the expected band structure, one can clearly see the existence of gap states in between the bands. In fact, for a system with $M$ barriers, every $M$-th state resides in a gap and corresponds to a state which is localised on either side of the box as a function of $\Delta$ (see Fig.~\ref{fig:EnergySpectrumTopologicalState}(b)). The chiral character of the individual states is indicated in Fig.~\ref{fig:EnergySpectrumTopologicalState}(a) by the different colours, with red corresponding to localisation on the left and blue corresponding to localisation on the right. However, for values of $\Delta$ at which the gap state energy has an extremum, the corresponding state loses its localised character and becomes extended over the whole box.

\begin{figure}[tb]
    \centering
    \includegraphics[width=\columnwidth]{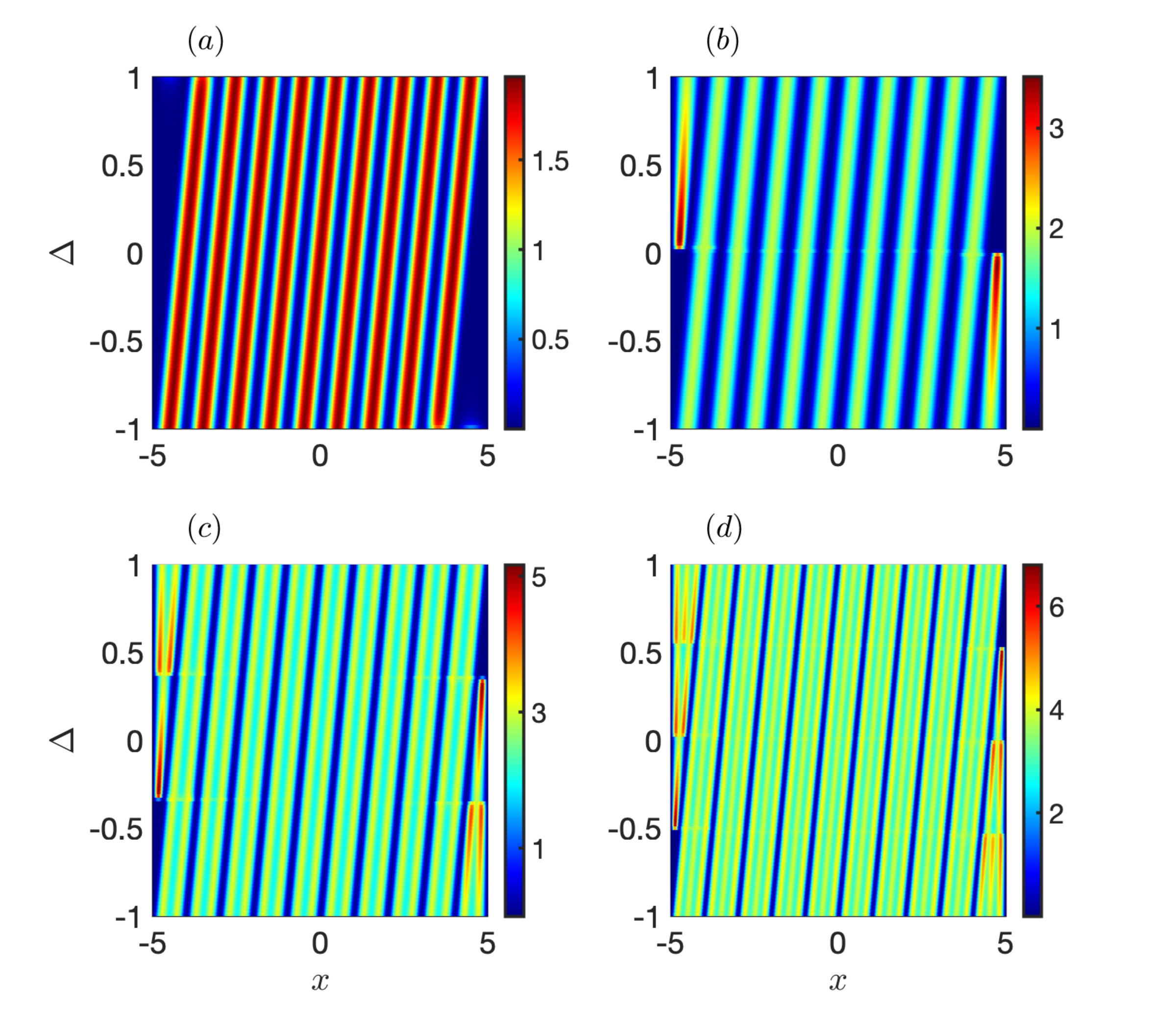}
    \caption{Integrated density of a Fermi gas at zero temperature in the AFKP potential with $10$ equidistant barriers of height $h=50$ in a box of size $L=10$. The  particle numbers are (a) $N=9$  (b) $N=10$, (c) $N=20$ and (d) $N=30$. Note the different color scales.
    }
    \label{fig:densitymany}
\end{figure}

While usually quantum many-body systems are hard to model exactly, ultracold Fermi gases at zero temperature do not exhibit s-wave scattering \cite{giorgini2008theory} and can therefore be described by only having the knowledge of the single particle eigenstates of the system Hamiltonian, $\psi_{k}(x)$. For a gas of $N$ atoms, the many-body wavefunction at zero temperature  can therefore be simply calculated from the Slater determinant as \cite{borland1970simple}
\begin{equation}
  \Psi(x_{1},x_{2},...,x_{N})=\frac{1}{\sqrt{N!}}\det_{k,j}^{N}[\psi_{k}(x_{j})],
  \label{eq:manybody}
\end{equation}   
with the total energy of the groundstate being given by $E_0=\sum_{k=1}^{N}{\varepsilon}_k$, where the ${\varepsilon}_k$ are the single particle eigenenergies.
 To check the effect the single particle edge states have on the many-body state of the gas, we will first look at the zero-temperature ground-state density for different numbers of particles in the system. For our system with $M=10$ barriers the integrated many-body density distribution, $N\int{|\Psi(x,x_{2},...,x_{N})|^2}dx_{2}...dx_{N}= \sum_{n=1}^{N}|\psi_{n}(x)|^2$
is shown in Fig.~\ref{fig:densitymany} for different particle numbers. Panel (a) show a system with $N=9$ fermions, and one can see that the density is fully delocalised, as it does not have any contributions from an edge state. Adding one more particle to the system (see panel (b)) requires one to include the single-particle edge state into the Slater determinant in Eq.~\eqref{eq:manybody}, which leads to an inhomogeneous many-body distribution with higher density at the edges of the box. Similarly, additional localized density peaks appear when including successively higher lying edge modes as shown for $N=20$ and $N=30$ in Figs.~\ref{fig:densitymany}(c) and (d). One can note that the number of peaks that appear in the density at the box edges corresponds to the number of occupied edge modes, and the different chiralities corresponding to the different values of $\Delta$ lead to them appear on the same or on opposite sides. The number of particles can, therefore, be seen as an additional variable that controls the chiral character of the many-body system. 

\section{Quenches and Overlaps} 
\label{sec:Fidelity}
While the stationary properties of topologically non-trivial systems are well studied in many settings \cite{bansil2016colloquium}, their dynamical properties remain less explored \cite{pastori2020signatures,Campbell14}. However, since in the AFKP model the chirality of each edge state can be altered by simply changing $\Delta$ (see Fig.~\ref{fig:Schematic}), it makes this model particularly suitable for studying out-of-equilibrium dynamics when quenching between parameter regimes with opposite chirality.
 Furthermore, at $\Delta=0$, every gap state reaches an extremum (see Fig.~\ref{fig:EnergySpectrumTopologicalState}(a)), which means that at this point all single particle gap states merge into the upper or lower energy bands and the corresponding eigenfunctions become delocalized. The system therefore also allows one to study quenches between fully delocalised systems and ones possessing an edge state. We will in the following explore the effect of edge states on the nonequilibrium dynamics of an ultracold Fermi gas (or its bosonic equivalent, the Tonks--Girardeau gas \cite{girardeau1960relationship}) by looking at the static and dynamic overlap of the states before and after the quench.
We will also look at the work probability distribution (WPD) to identify the dominating excitations after the quench and their dependence on the presence or absence of an edge mode near to the Fermi level.

\subsubsection{Static Overlap}
 
The overlap between the groundstates of two different Hamiltonians give the transition probability when driving one state to the other using a sudden quench, and is therefore related to the amount of excitations that are created when the system is driven out of equilibrium. Here we focus on sudden quenches of the lattice shift parameter $\Delta_1\rightarrow\Delta_2$, for which the static overlap of the respective groundstates $\Psi_{\Delta_1}$ and $\Psi_{\Delta_2}$ is given by
\begin{equation}
    \nu=\braket{\Psi_{\Delta_1}(x_1,x_2,\dots,x_N)|\Psi_{\Delta_2}(x_1,x_2,\dots,x_N)}.  
    \label{eq:static}
\end{equation} 
In Anderson's original work it was shown that the overlap between two many-body wavefunctions continuously decreases with increasing system size as $\nu\propto N^{-\frac{\alpha}{2}}$  \cite{anderson1967}. Here $\alpha$ is related to the perturbation strength, showing that the overlap can vanish even for infinitesimal perturbations for large enough particle numbers $N$. While calculating the overlap in Eq.~\eqref{eq:static} using full many-body wavefunctions is difficult and time consuming, for non-interacting fermions it can be directly calculated from the single particle eigenstates $\ket{\psi_{\Delta_1,n}}$ and $\ket{\psi_{\Delta_2,k}}$ of the respective Hamiltonians with $\Delta_{1}$ and $\Delta_{2}$ as
\begin{equation} 
    \nu=\det [B],
\end{equation}
where the matrix $B$ is defined by the elements $B_{k,n}=\braket{\psi_{\Delta_2,k}|\psi_{\Delta_1,n}}$ for $k,n=1,\dots,N$ \cite{anderson1967}. While the OC has been extensively investigated in condensed matter and cold atomic systems for its importance in understanding quantum many-body systems 
\cite{mahan2013many,goold2011orthogonality,sindona2013orthogonality,knap2012time,Campbell14} and quantum phase transitions \cite{Lelas_2012,Campbell_2016}, 
it is interesting to explore how the presence of chiral edge states affects the many-body overlap.

\begin{figure}[tb]
     \centering
    \includegraphics[width=\columnwidth]
    {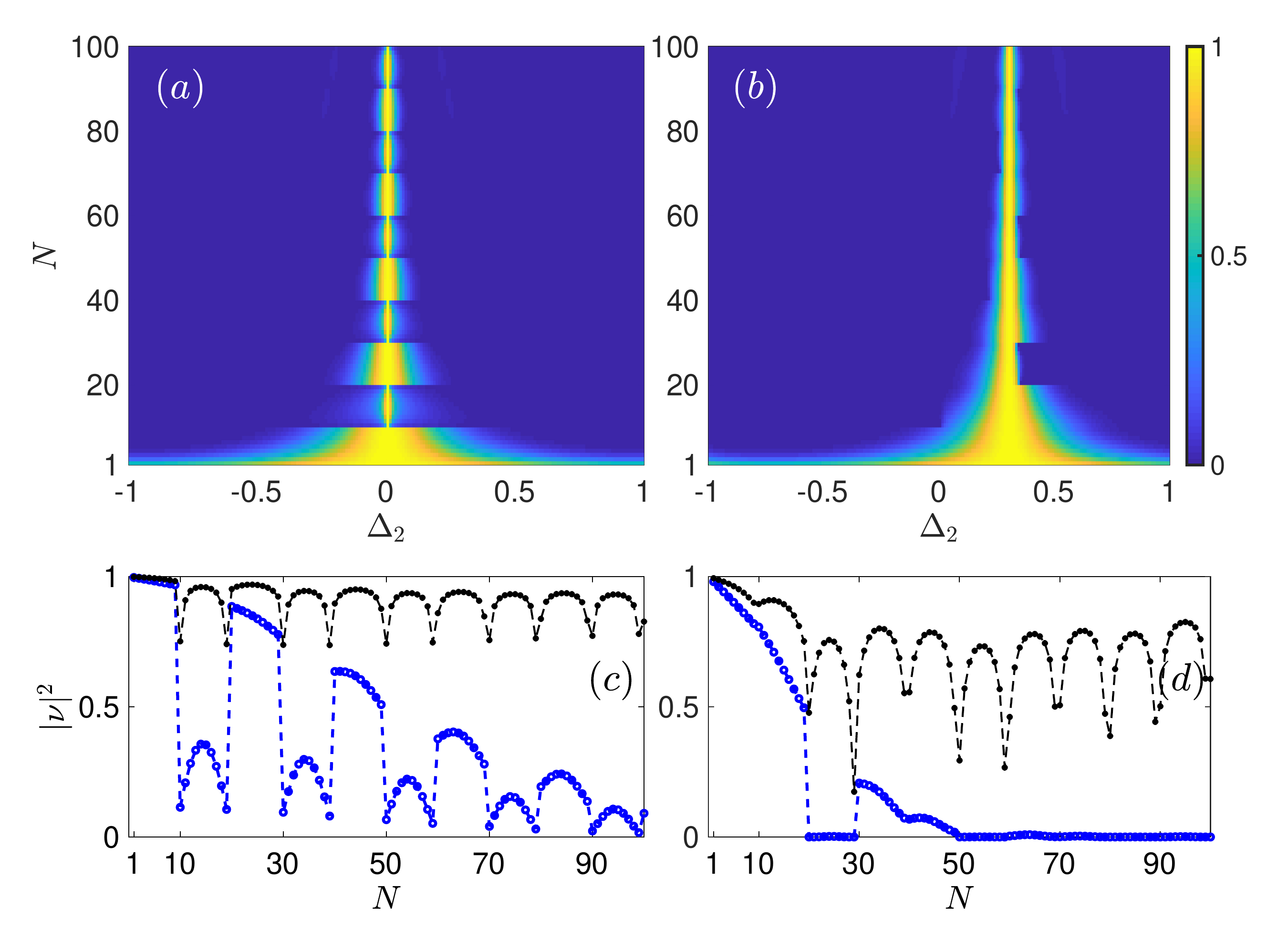}
    \caption{(a) Transition probabilities $|\nu|^2$ for  (a) $\Delta_1=0$ and (b) $\Delta_1=0.3$ and all other $\Delta_2$ settings for a lattice of barrier height $h=50$. (c) The blue dots corresponds to a cut through panel (a) at $\Delta_2=0.04$ and in (d) the blue dots corresponds to a cut through panel (b) at $\Delta_2=0.4$. The black dots in both panels corresponds to the same values for a lattice with $h=5$.  
    }
    \label{fig:staticoverlap}
\end{figure}

For this we first compare a system with $\Delta_1=0$, for which no edge state is present, with groundstates of potentials with $\Delta_2\neq 0$. The corresponding transition probabilities $|\nu|^2$ are shown in Fig.~\ref{fig:staticoverlap}(a), and one can see an overall decay with increasing particle number for all values of $\Delta_2$, in agreement with the principle of the orthogonality catastrophe \cite{anderson1967}. However, one can also see that for specific ranges of particle numbers, there are discontinuous jumps in the transition probability where it abruptly decreases. These ranges are of finite size and Fig.~\ref{fig:staticoverlap}(c) shows  the specific case of $\Delta_{2}=0.04$ (blue line). One can see that the overlap deviates from the power-law decay in the intervals from $N=10$ to $N=19$, from $N=30$ to $N=39$ and in corresponding intervals for higher values of $N$. 

The reason for this unusual behaviour is the presence of the chiral 
edge states inside each of the multiple bandgaps in the energy spectrum. While initially, for $\Delta_1=0$, all eigenstates of the system are delocalised for all particle numbers, after the quench, for $\Delta_2=0.04$, the system possesses chiral edge states and in our example the first one of these is the 10$^\text{th}$ single particle eigenstate.  
Therefore, if the static overlap is calculated between the two many-body groundstates corresponding to $\Delta_1=0$  and $\Delta_2=0.04$ in a system of up to $9$ particles, all single particle overlaps on the diagonal of $B$ are calculated using extended states which possess a rather large overlap. 
However, by considering a system with $N=10$ particles, the highest-lying state that needs to be taken into account is given by a localized edge state for $\Delta_2=0.04$ which therefore has significantly reduced overlap with the initial extended states. This leads to a sudden decrease in the many-body transition probabilities, which is also clearly visible for all values of $\Delta_2$ in Fig.~\ref{fig:staticoverlap}(a). When the number of particles in the gas increases to $N=20$, a second edge state, which for $\Delta_2=0.04$ is localised on the opposite edge of the box (see Fig.~\ref{fig:EnergySpectrumTopologicalState}(a)) appears and  the symmetry of the total density of the system is partially restored. This, in turn, significantly increases the many-body overlap again. Since at $\Delta=0.04$ subsequent edge states also have opposite chirality, the same behaviour manifests for systems with between $N=21$ and $N=40$ particles and each consequent interval. 

Let us next discuss the situation where the initial set of eigenstates is not entirely delocalised, but already contains edge states.
For this we consider the overlap between the groundstates for  $\Delta_1=0.3$ with states at all other values $\Delta_2$ and the resulting transition probabilities can be seen in Fig.~\ref{fig:staticoverlap}(b). While again an overall decay for increasing particle numbers appears, one can also immediately note that the gaps are no longer symmetric as in the previous case and are located in different particle intervals. 
To interpret this, let us consider the static overlap between $\Delta_1=0.3$ and  $\Delta_2=0.4$ which is shown in detail in Fig.~\ref{fig:staticoverlap}(d). One can see that for small particle numbers an initial decay is present, with a  change in slope at $N=11$, however there is no sudden decrease of the overlap.  This can be understood by realising that for $9$ particles all states are delocalised, and the edge state occupied by the $10^\text{th}$ particle is localised on the left hand side of the box for lattice shifts of both $\Delta_1=0.3$ and  $\Delta_2=0.4$.
The overlap between the corresponding many-body states therefore stays large and finite, as their structure is the same.
When increasing the number of particles up to 20, and therefore occupying the next higher-lying edge mode, the system with $\Delta_1=0.3$ has its second edge state localised on the right of the box, while the system with $\Delta_2=0.4$ has its second edge state localised at the left of the box (see the dashed lines in Fig.~\ref{fig:EnergySpectrumTopologicalState}(a)). The many-body density for $\Delta_2=0.4$ is therefore more localized on the left edge of the box compared to the more homogeneous one of $\Delta_1=0.3$, leading to a rapid decrease of the overlap (see the line in blue in Fig.~\ref{fig:staticoverlap}(d)). When occupying the third edge state using $N=30$ particles, the occupation of an edge state for $\Delta_1=0.3$ at the left of the box and on the right for $\Delta_2=0.4$ delocalises the two many-body densities again, which results in the visible jump of transition probability to a higher value. This pattern continues for even larger particle numbers, depending on the occupation of successive single particle edge states and their chirality.

To explore the influence of the height of the lattice, and therefore the strength of the localisation of the chiral edge states, we also show in Figs.~\ref{fig:staticoverlap}(c) and (d) the overlaps for a system with lattice height $h=5$ (black dots). In such a system the edge states are only weakly localized at the edges of the box and have significant spread into adjacent wells. This diminishes the effect of chirality and the overlaps can be seen to not display the gaps across certain particle intervals. Instead, the overlap decays overall slowly and is punctuated with sharp minima which appear whenever a the Fermi energy is at the top of a band. This is similar to what is seen in the pinning transition of a Tonks--Girardeau gas in an optical lattice potential \cite{Lelas_2012}, and the strong influences of the chiral edges states are lost.

\subsubsection{Dynamic Overlap}

To study the impact of the edge states on the orthogonality catastrophe as it would be observed in an experiment \cite{MarkoCetina}, we calculate the dynamical overlap after a sudden quench from one lattice shift to another
\begin{equation}
    \nu(t)=\braket{\Psi_{\Delta_1}|e^{iH_1t}e^{-iH_2t}|\Psi_{\Delta_1}},
    \label{eq:dynamicoverlap}
\end{equation} 
where $H_1$ is the Hamiltonian of the initial potential with a lattice shift of $\Delta_1$, and $H_2$ is the final Hamiltonian  with a lattice shift of $\Delta_2$. The state $\Psi_{\Delta_1}$ is the initial state and the groundstate of $H_1$. As before, for systems of non-interacting fermions the dynamical overlap can be rewritten as 
\begin{equation}
   \nu(t)=\det[A(t)],
\end{equation}
where the elements of the matrix $A(t)$ are given in terms of single particle overlaps
\begin{equation} 
    A_{k,n}(t)=\bra{\psi_{\Delta_1,k}} e^{i\mathcal{H}_1t} e^{-i\mathcal{H}_2t} \ket{\psi_{\Delta_1,n}}.
    \end{equation}
Rewriting this expression in terms of the single particle states $\psi_{\Delta_2,l}$ of $\mathcal{H}_2$ gives 
\begin{equation} 
    A_{k,n}(t)=\sum_{l=1}^{\infty}\braket{\psi_{\Delta_1,k}|\psi_{\Delta_2,l}}\braket{\psi_{\Delta_2,l}|\psi_{\Delta_1,n}}e^{-i({\varepsilon'_{l}}-\varepsilon_{k})t},
\end{equation}
where ${\varepsilon}_k$ and $\varepsilon_l'$ are the eigenenergies of $\mathcal{H}_{1}$ and $\mathcal{H}_{2}$ respectively. Since the quench dynamics leads to excitations in the system, the results can be expected to be quite different from the static case discussed above.

\begin{figure}[tb]
    \centering
    \includegraphics[width=\columnwidth]{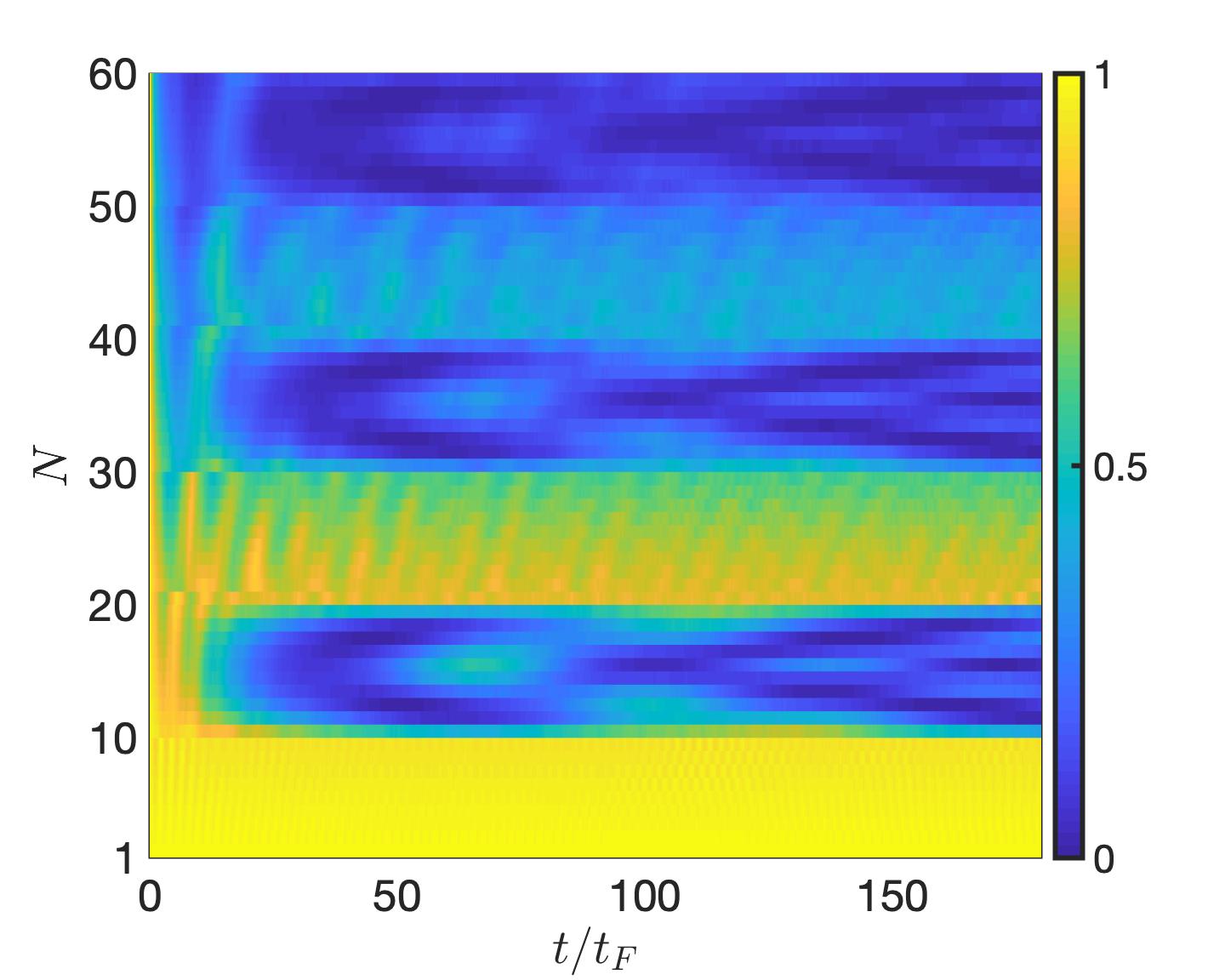}

\caption{The survival probability for a quench from $\Delta_1=0$ to $\Delta_2=0.04$ for different numbers of particles $N$ for a lattice of barrier height $h=50$. 
    }
    \label{fig:dynamicoverlap}
\end{figure}

The survival probability $|\nu(t)|^{2}$  allows one to quantify the non-equilibrium dynamics after the quench and it is shown in Fig.~\ref{fig:dynamicoverlap} as a function of the particle number $N$.  Here we consider a quench from $\Delta_1=0$ to  $\Delta_2=0.04$ (cf.~Fig.~\ref{fig:staticoverlap}(c)) and the time is re-scaled in units of the Fermi time $t_F=1/E_{F}$, where $E_{F}=\varepsilon'_{N}$ is the Fermi energy of $\mathcal{H}_2$. While the detailed dynamics can be seen to be rather complex, the dominating effects of the chiral edge states are still visible in the ensuing magnitudes of the survival probabilities: almost no decay is visible as long as $N\le 9$, but the survival probability decreases rather quickly in the interval of particle numbers where only one edge mode is occupied and increases again with the inclusion of the next edge mode once the system consists of 20 particles. This is consistent with the results in the static situation and the explanation for the appearance of the different zones in Fig.~\ref{fig:dynamicoverlap} is therefore the same for the results of the transition probability. Similar dynamics for fillings that differ by 20 particles is due to the repeating single-particle energy curvature pattern in the vicinity of $\Delta_1$ and $\Delta_2$ as seen in Fig.~\ref{fig:EnergySpectrumTopologicalState}(a).

\begin{figure*}[tb]
    \centering
    \includegraphics[width=2\columnwidth]{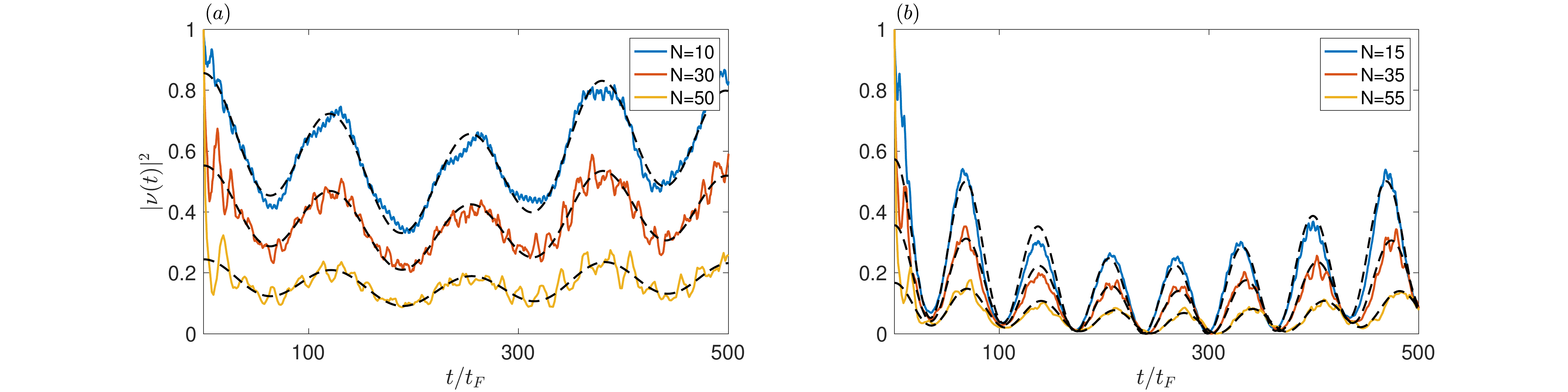}
    
    \caption{Survival probability $|\nu(t)|^2$ for a quench from $\Delta=0$ to $\Delta=0.04$ for (a) $N = 10$, $30$ and $50$ particle fillings and (b) for $N = 15$, $35$ and $55$. The dashed black lines correspond to the survival probabilities constructed from excitations with the highest contribution in the work probability distribution described by Eq.~\eqref{eq:wpdoverlap}. 
    }
    
    \label{fig:fidelitycomp}
\end{figure*}

To better understand the structure of the survival probability, we look at its time evolution for systems with $N=10$, $30$ and $50$ particles in Fig.~\ref{fig:fidelitycomp}(a). For the selected fillings an odd number of alternating chiral edge states contribute to the post-quench Slater determinant leading to similar dynamics for all of them. As predicted by the OC, systems with larger particle numbers have a lower survival probability. 
However, a different behaviour is visible for systems with particle numbers $N=15$, $35$ and $55$ as shown in Fig.~\ref{fig:fidelitycomp}(b), with the survival probability quickly decaying and even reaching zero at certain times. The presence of dynamical orthogonality in these cases is unexpected when comparing with the static overlap in Fig.~\ref{fig:staticoverlap}(c) which shows that the dynamical excitation of the systems is unlikely to leave the system in the post-quench groundstate, and therefore stimulate  more irreverisble dynamics. Since states in the band above each edge state are de-localised they have larger transition probabilities from the uniform pre-quench eigenstates and should therefore have a larger survival probability. This reduces the effect of chirality on the system.

\subsubsection{Work probability distribution} 
\begin{figure}[tb]
    \centering
    \includegraphics[width=\columnwidth]
    {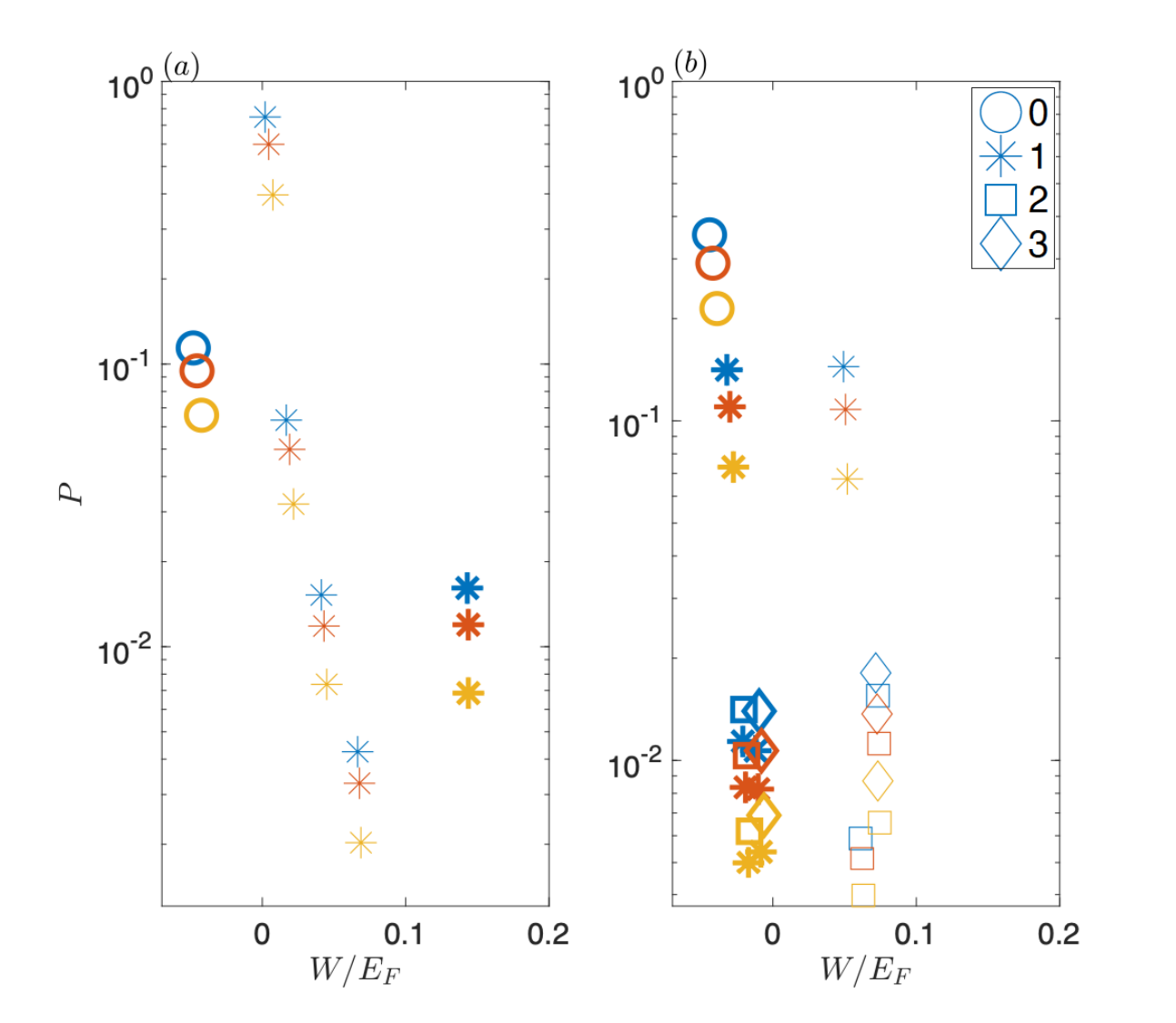}
   \caption{Work probability distribution for a quench from $\Delta_1=0$ to $\Delta_2=0.04$ for systems with (a) $10$ (blue), $30$ (orange) and $50$ (yellow) particles; (b) $15$ (blue), $35$ (orange) and $55$ (yellow) particles. Bold symbols show the excitations that possess chiral edge modes where their total density is localized on one side of the trap. The energy is scaled in units of the respective Fermi energies $E_F$.}
  \label{fig:excitation_identical}
\end{figure}

To understand the role of the edge states in the dynamics of the survival probability we must look at the spectral decomposition through the work probability distribution (WPD), which provides information on how excitations contribute to the arising dynamics \cite{campisi2011colloquium, PhysRevX.4.031029, zhang2022work}. The WPD is given by the Fourier transform of the dynamical overlap $P(W)=\int dt\; e^{-iWt} \nu(t)$ and may be written as
\begin{equation}
     P(W)=\sum_{m}|\langle\Psi|\Phi_{m}\rangle
    |^{2} \delta(W-(E_{m}^{'}-E_{0}))\,,
\end{equation}
where $\Psi$ is the pre-quench many-body groundstate with energy $E_{0}$ and the $|\Phi_m\rangle$ are the many-body eigenstates of the quenched Hamiltonian with total energy $E_{m}^{'}$. This expression therefore quantifies the excitations in the Hilbert space of the final Hamiltonian. The sum runs over all possible excitation configurations and different excitations can be identified by comparing them to the initial Fermi sea. In a system of $N$ particles, a single particle excitation corresponds to the lowest $N-1$ states being occupied and one particle being in a state above the Fermi level. An excitation of this type is indicated by $*$ in Fig.~\ref{fig:excitation_identical}. If the lowest $N-2$ states are occupied and two particles occupy states above the $N-1$ state, a two-particle excitation exists (indicated by a $\Box$), with higher excitations defined in a similar way \cite{campisi2011colloquium}. If the system is excited to the new post-quench ground state, this is marked using circles in Fig.~\ref{fig:excitation_identical}.

The WPD for systems with particle numbers $N=10$, $30$ and $50$, for which the highest occupied states pre-quench are edge-states, is shown in Fig.~\ref{fig:excitation_identical}(a). In each case, three excitations have probability values $P_m \equiv |\langle\Psi|\Phi_{m}\rangle|^{2}$ significantly larger than the rest, which means that they can be expected to provide the largest contribution to the dynamical evolution of the overlap. If we reconstruct the survival probability
\begin{equation}
     |\nu(t)|^2=\left|\sum_{m} P_m e^{i (E'_m-E_0)t}\right|^2, 
 \label{eq:wpdoverlap}
\end{equation}
using these three largest values of $P_m$ we indeed recover a good agreement with the exact dynamics seen in Fig.~\ref{fig:fidelitycomp}(a), where the obtained approximate dynamics are depicted by dashed black lines. For the system with $N=10$ the agreement is the closest since the three largest probabilities sum up to $\sum_{m=1}^3 P_m = 0.92$, meaning that all higher energy excitations have a minimal contribution. Increasing the particle number decreases the value of this sum as other excitations become important, as evident from the emerging high frequency modes in the exact calculations for the $N=30$ and $N=50$ cases. Nevertheless, the overall long-time behaviour is captured accurately by the approximation. 

The most significant contributing mode in the WPD in each case is a low-energy single-particle excitation from the Fermi surface to the next available single particle state at the bottom of the next energy band. As the energies of these two states are very close (see Fig.~\ref{fig:EnergySpectrumTopologicalState}) even a small amplitude quench is enough to excite the system. Importantly, neither the initial state nor this post-quench state contain an edge mode and so both their densities are uniform and the overlap is large. We highlight chiral excitations, in which total density is localized on one side of the trap, by the bold symbols in Fig.~\ref{fig:excitation_identical}(a). The nearest post-quench state that is chiral is the groundstate, however its probability is small compared to the largest single particle excitation and therefore it does not have a large effect on the dynamics. Consequently, the survival probability in Fig.~\ref{fig:fidelitycomp}(a) does not decay to zero as the main contributions are from delocalized states and the edge states do not play a significant role. 

If the Fermi level is within the middle of the band, as  for the $N=15$, $35$ and $55$ particle fillings, the dominant term in the WPD is the transition to the post-quench ground state (see Fig.~\ref{fig:excitation_identical}(b)), however due to the close energy level spacing within the band, higher particle number excitations are relevant as well (see $\Box$ and $\Diamond$ symbols). To reproduce the approximate dynamics seen in Fig.~\ref{fig:fidelitycomp}(b), one therefore needs to consider the post-quench states with the five highest probabilities, and the increased number of involved excitations invariably leads to a visible decay of the survival probability. Indeed, in this case, the most probable post-quench state is the groundstate, which is a chiral edge state and therefore is localized on the left side of the trap. In this case, there is also an increased number of higher energy excitations to localized chiral edge states as highlighted by the bold symbols in Fig.~\ref{fig:excitation_identical}(b)). This results in a rapid change of the total density as the chiral edge modes are excited by the quench and consequently the survival probability decays quickly towards zero (see Fig.~\ref{fig:fidelitycomp}(b)).  
Overall, the WPD clearly identifies the excitation structure responsible for the dynamics of the survival probability with increasing particle number, showing that it is determined by the Fermi level of the initial state and its proximity to the closest edge mode. 

\section{Conclusion}
\label{Sec:conclusions}

In this work, we have investigated the role of chiral single-particle edge states in the quench dynamics of a quantum many-body system. Using a gas of ultracold fermions at zero temperature and confined in an AFKP potential, we have analyzed the transition probability, survival probability, and work probability distribution following a sudden change in the lattice shift parameter.

Our findings reveal a significant departure from the conventional monotonic decay of the overlap, as predicted by the orthogonality catastrophe, with increasing particle number. These deviations arise from the presence of chiral single-particle edge modes in the system, which emerge as particle levels are filled. Consequently, the particle number serves as an additional parameter influencing the chiral nature of the many-body system. The work probability distribution further demonstrates that, for specific particle fillings, a quench can selectively excite particles in and out of certain edge modes, leading to intricate dynamics in the dynamic overlap. The ability of the quench dynamics to probe chirality is highly sensitive to the choice of the initial state, particularly the position of the Fermi energy relative to the nearest single-particle edge mode.

Overall, our results highlight how chiral single-particle edge states in a topologically nontrivial model shape the nonequilibrium dynamics of fermionic many-body systems. This work therefore provides a deeper understanding of quantum many-body behavior under a quench scenario. A promising avenue for future research is to examine how chirality influences nonequilibrium dynamics in many-body systems with finite interactions, where the single-particle description used here no longer applies. Suitable candidates for such studies include bosons interacting via $s$-wave scattering or multi-component fermionic systems with $SU(N)$ symmetry.

\section*{Acknowledgements}
This work was supported by the Okinawa Institute of Science and Technology Graduate University. The authors are grateful for the Scientific Computing and Data Analysis (SCDA) section of the Research Support Division at OIST. T.F. acknowledges support from JSPS KAKENHI Grant No. JP23K03290. T.F. and T.B. are also supported by JST Grant No. JPMJPF2221 and JSPS Bilateral Program No. JPJSBP120227414.

\bibliography{TopoQuench}

\end{document}